\documentstyle[epsfig]{article}
\begin{document}

\def\scri{\unitlength=1.00mm
\thinlines
\begin{picture}(3.5,2.5)(3,3.8)
\put(4.9,5.12){\makebox(0,0)[cc]{$\cal J$}}
\bezier{20}(6.27,5.87)(3.93,4.60)(4.23,5.73)
\end{picture}}

\begin{center}

{\bf Some Examples of Trapped Surfaces}

\vspace{8mm}


Ingemar Bengtsson

\

{\sl Fysikum, Stockholms Universitet} 

{\sl 106 91 Stockholm, Sweden}

\end{center}

\

\noindent We present some simple pen and paper examples of trapped surfaces in 
order to help visualising this key concept of the theory of gravitational 
collapse. We collect these examples from time-symmetric initial data, 2+1 
dimensions, collapsing null shells, and the Vaidya solution.

\vspace{1cm}

{\bf 1. Introduction}

\

\noindent Provided an appropriate positivity property holds the existence of a closed 
trapped surface has dramatic consequences for the future evolution of spacetime. 
In numerical relativity trapped surfaces provide---assuming that a cosmic censor is 
active---the practical means by which black holes are recognized, and indeed effective 
algorithms to spot trapped surfaces in a given initial data set have been developed 
\cite{Thornburg}. There are also theorems in mathematical relativity that 
tell us much about the existence of marginally trapped surfaces 
\cite{Hawking, RPAC, Kriele, MS, Eichmair, Andersson, Carrasco},  their 
``evolution'' into marginally trapped tubes foliated by such surfaces \cite{Ashtekar, 
Lars}, and their formation from initial data sets that are free of them 
\cite{Big}. A possible application is to quasi-local definitions of black 
holes \cite{Tipler, Hayward, AK, Booth}: the numerical algorithms 
are not called ``horizon finders'' for nothing. Other chapters in this 
book provide the details. The modest aim here is 
a supplementary one: we will collect some illustrative 
examples of trapped surfaces that one can obtain with pen and paper only.

After some preliminaries in section 2, we discuss time-symmetric initial data of 
the Einstein equations in section 3. Early studies initiated by Wheeler and 
pursued by Brill and Lindquist \cite{BL} and many others \cite{Misner, Gibbons, 
Ca, Bi} led to the construction 
of initial data containing an arbitrary number of Einstein-Rosen bridges to 
new asymptotic regions, and to interesting observations on the minimal surfaces 
that appear there. Although the term was not used---because the notion of trapped 
surfaces had not yet been introduced---these surfaces are in fact marginally 
trapped in the resulting spacetime. In section 4 we restrict ourselves to a 
2+1 dimensional toy model of gravity (with a negative cosmological constant) 
\cite{BTZ, brapapper, Dieter, Steif}, where a complete description of all 
marginally trapped ``surfaces'' in spacetime 
can be had \cite{Brill}. In this toy model we will 
also see how the trapped surfaces ``jump'' when we throw lumps of matter into 
a black hole \cite{Emma}, and---by increasing the dimension again---what a 
dynamical horizon can look like in the vacuum case \cite{Peldan}.   

In 3+1 dimensions trapped surfaces can be produced in a controlled manner by 
sending convex shells of incoherent radiation into flat spacetime. This idea 
was proposed by Penrose \cite{Penrose} as a way to test cosmic censorship; it 
has been much studied \cite{Tod, Gary, Pelath} and will be discussed in section 5. 
Following, say, a marginally trapped tube into the future is a more difficult 
matter, but in section 6 we say more about the spherically symmetric Vaidya 
solution \cite{Papa, BenSen, BS, ABS, Nielsen, Ben-Dov}. This is possible because 
spherical symmetry prevents gravitational radiation from appearing behind the 
shell. 

For natural reasons there will be a certain emphasis on trapped surfaces I 
have encountered myself---with apologies to authors who feel, perhaps rightly 
so, that their own examples are more interesting.

\

{\bf 2. Preliminaries}

\

\noindent We are interested in the way a submanifold is embedded into some larger 
space, so for this purpose we decompose the 
tangent space at a point of the submanifold into a tangential and a normal part. 
If $X$ and $Y$ are vector fields belonging to the former the Weingarten or 
shape tensor $K$ will produce a vector belonging to the latter according to its 
definition 

\begin{equation} K(X,Y) = - (\nabla_XY)^\perp \ . \end{equation}

\noindent Here $\nabla_X$ is the usual covariant derivative along the 
tangential direction $X$. 
For an actual computation we may assume that the submanifold is given 
in parametric form, $x^a = x^a(u)$, and we find that the Weingarten tensor 
contracted into a normal vector $k^a$ is 

\begin{equation} K_{ij}(k) = - k_a\frac{\partial^2 x^a}{\partial u^i \partial u^j} 
- k_a\Gamma_{bc}^{\ \ a}\frac{\partial x^b}{\partial u^i}\frac{\partial x^c}
{\partial u^j} \ . \label{friendly} \end{equation}

\noindent In the examples below such a calculation is typically either 
straightforward but tedious, or hideously complicated. The details will be 
relegated to the list of references.  

If the submanifold has codimension 1 the normal vector is unique, and 
the Weingarten tensor is referred to as the second fundamental form---but trapped 
surfaces have codimension 2 by definition. In any case the first fundamental form 
$\gamma_{ij}$ is the metric induced on the submanifold, and the mean curvature 
vector is defined by 

\begin{equation} H^a = \gamma^{ij}K_{ij}^a \ . \end{equation}

\noindent In Riemannian spaces it can be shown that the surface is 
minimal---meaning that its area 
always increases if the surface is subject to variations of sufficiently small 
compact support---if and only if its mean curvature vector vanishes. For 
hypersurfaces the mean curvature vector contracted into the unique normal 
vector is called the mean curvature, denoted $K$.  

As part of its definition a trapped surface is a spacelike surface of 
codimension 2, in a Lorentzian spacetime. It follows that any normal vector 
can be expressed as a linear combination of two future directed null vectors, 
normalised by 

\begin{equation} k_+\cdot k_- = -2 \ . \end{equation}

\noindent As the notation suggests, one of these vectors ($k_+$) will be directed 
``outwards'' and the other ``inwards''. The set of all such vectors will 
give rise to one outgoing and one ingoing null congruence, and the surface is 
said to be trapped if the cross sections of both congruences decrease in area 
as they leave the surface. Whether this is so can be read off from the mean 
curvature vector, which is 

\begin{equation} H^a = - \theta_+k_-^a - \theta_-k^a_+ \ . \end{equation}

\noindent The surface is trapped if both the null expansions $\theta_\pm$ are 
negative, meaning that the mean curvature vector is timelike and future 
directed. The surface 
is marginally trapped if the outer expansion $\theta_+ = 0$ and the inner 
expansion $\theta_- \leq 0$. 

Actually, in the algorithms that look for marginally trapped surfaces, only 
the outwards expansion is controlled \cite{Thornburg}. Hence they produce 
marginally outer trapped surfaces, where the meaning of ``outer'' is provided 
by some spacelike slice in which the surface is embedded. It turns out that 
the main body of 
mathematical relativity theorems concerns marginally outer trapped surfaces, 
and that such surfaces suffice to prove singularity theorems \cite{Lars}. But 
we will concentrate on genuinely trapped and marginally trapped surfaces (with 
$\theta_- \leq 0$) here, partly because the marginally trapped tubes swept 
out by a sequence of such surfaces have particularly interesting properties 
\cite{Hayward, AK}.  

In 2+1 dimensional Minkowski space, which is helpful for visualisation, a 
codimension 2 submanifold is simply a curve. Let it be spacelike. It is future 
trapped if its Frenet 
normal vector is timelike and future pointing, and untrapped if the normal 
vector is spacelike. A marginally outer trapped curve is a curve in some 
null plane, and its inner expansion is negative if and only if that curve 
is concave. It will be observed that only the untrapped curves can be 
closed. From now on, when we talk of a trapped surface we 
often assume that it 
is closed---since otherwise its existence has no particular consequences. 
In this sense there are no closed trapped surfaces in Minkowski space. 
If space has the topology of a cylinder we find that there are closed 
marginally outer trapped curves in suitable null planes, but still none 
with $\theta_- < 0$ everywhere. 

\begin{figure}
        \centerline{ \hbox{
                \epsfig{figure=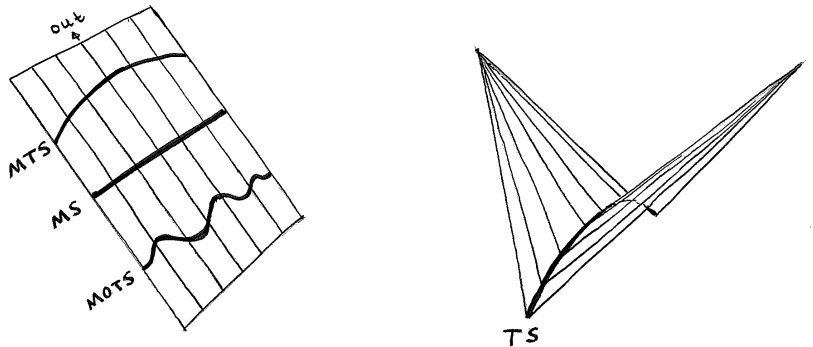,width=110mm}}}
        \caption{\small A null plane containing marginally trapped, 
        minimal, and marginally outer trapped curves. Note that any 
        spacelike plane crossing the null plane will contain a curve 
        of the latter type. A locally trapped curve (a piece of a 
        hyperbola bending down in time) is on the right.}
        \label{fig:trappex5}
\end{figure}

The Riemannian cousins of marginally trapped surfaces are the minimal surfaces, 
and many beautiful examples have been published. Based on Plateau's experiments 
with soap films spanned by thin wires \cite{Plateau}, it was conjectured that 
one can find a minimal surface with any given one dimensional boundary. This 
was proved with mathematical precision some eighty years later \cite{Blaine}. But 
the boundary is essential---closed minimal surfaces do not exist in 
Euclidean space. A 
way to see this is to observe that the equation obeyed by the functions 
$x^a(u)$ when the first variation of the area functional vanishes is the 
Laplace equation. 

Thus in flat space a minimal surface needs a boundary, but we can do better 
if we are inside the 3-sphere. Indeed the solution of 
the Plateau problem, together with a scheme using reflections, can then be 
used to produce closed minimal surfaces of any genus \cite{Lawson}. For genus 
0 and 1 it is easy to see what these minimal surfaces are. The equator of 
the 3-sphere is a minimal 2-sphere, and the maximal torus that appears in 
the Hopf fibration is a minimal 2-torus. Like the geodesics---the great 
circles on the sphere---they are in a sense ``maximal'' rather than 
minimal, because there exists a global deformation that decreases their 
area. The technical term for such minimal surfaces is ``unstable''.

Although they are not based on any variational principle, the notion of
stability can be generalised so as to apply to marginally outer trapped surfaces 
\cite{RPAC, Lars}. Moreover a Plateau problem can be formulated and solved for 
them \cite{Eichmair}.

A totally geodesic submanifold is a submanifold such that the entire Weingarten 
tensor vanishes. This means that a geodesic starting out tangential 
to the submanifold will stay in the surface, where it defines a geodesic with 
respect to the intrinsic metric as well. Unless their dimension is equal to 
one---a geodesic---such submanifolds typically do not exist, but if the embedding 
space has a large amount of symmetry they may. In particular, the fixed point set 
of an isometry is always totally geodesic. (This is easy to see. Consider 
a vector tangent to the surface at some point. This defines a unique 
geodesic in the embedding space. If this geodesic were able to move out of 
the surface a part of it would be moved by the isometry, while its starting 
point and its initial tangent vector would not be. This contradicts 
the uniqueness of the geodesic.) In general relativity a totally geodesic 
spacelike hypersurface occurs whenever time-symmetric initial data exist, 
and we can then rely on the useful fact that a minimal surface in such an 
initial data hypersurface is a marginally trapped surface in the resulting 
spacetime. 

\

{\bf 3. Time-symmetric initial data}

\

\noindent The initial data for Einstein's equations are given by the first and 
second fundamental forms (two symmetric tensor fields on some space, one of them 
being its metric), subject to suitable constraints. If we take the 
second fundamental form to vanish---so that we will have a totally geodesic 
hypersurface in the spacetime to be constructed---the constraint on the 
first fundamental form take the simple form

\begin{equation} R^{(3)} = 0 \ . \label{R} \end{equation}

\noindent Its curvature scalar vanishes. 
This assumes vacuum, with vanishing cosmological constant. We may 
simplify matters further by assuming that the intrinsic metric takes the form 

\begin{equation} ds^2 = \omega^4(dx^2 + dy^2 + dz^2)
\ . \end{equation}

\noindent This was referred to as ``geometrostatics'' in the early references 
\cite{BL, Misner}. The reason for the unusual exponent on the conformal 
factor is that eq. (\ref{R}) now takes the form 

\begin{equation} \triangle \omega = 0 \ , \end{equation}

\noindent where $\triangle$ is the flat space Laplacian. From electrostatics 
we know how to solve this. A suitable solution is 

\begin{equation} \omega 
= 1 + \sum_{i=1}^N\frac{e_i}{r_i} \ , \end{equation}

\noindent where the source points can be placed at $N$ arbitrary positions,  
and $r_i$ is the Euclidean distance to the $i$th source point. 

If there is only 
one source point, placed at the origin, and if we use 
spherical polar coordinates $(r,\theta, \phi )$, this is the $t=0$ slice of the 
Schwarzschild solution in isotropic coordinates. Its mass $M = 2e$. This 
geometry admits a discrete isometry under 

\begin{equation} r \rightarrow r' = \frac{e^2}{r} \ . \end{equation}

\noindent From this we learn two things. First, that the source points in 
the solution do not represent singular points in the geometry, they represent 
the spatial infinities in $N$ (out of $N+1$) asymptotic regions. Second, when 
$N = 1$ the sphere at $r = e = M/2$ is a fixed point set of an isometry, hence it 
is totally geodesic and by implication a minimal surface. When the geometry 
evolves it becomes the bifurcation sphere on the event horizon.  

\begin{figure}
\centerline{ \hbox{
                \epsfig{figure=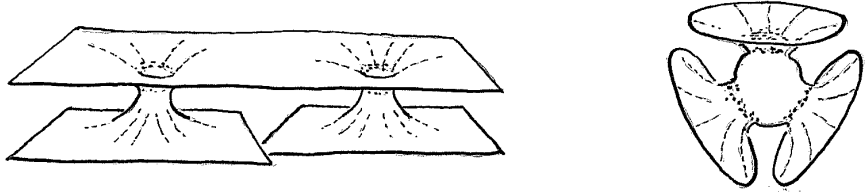,width=120mm}}}
\caption{\small{Brill-Lindquist initial data: When the throats 
are far apart there are only two minimal surfaces. In the symmetric 
case shown to the right there are obviously three.}} 
\label{fig:trappex1}
\end{figure}

The case when there are more than two asymptotic regions (when $N > 1$) is 
made intuitively clear in the illustrations provided by Brill and Lindquist 
\cite{BL}. See Fig. \ref{fig:trappex1}. There will be $N$ throats leading 
from the original asymptotically 
flat sheet to an additional $N$ such sheets. In each throat we will find 
a minimal 2-sphere, but it will no longer be a round sphere---rather it 
will be distorted by its interaction with the others. If the throats are 
well separated this is the end of the story, but if two throats come 
sufficiently close together an additional minimal sphere will appear and 
surround the two minimal spheres in the throats. 

To see how this works, let us choose $N = 2$ and $e_1 = e_2$. The conformal 
factor is  

\begin{equation} \omega + \frac{e}{r_1} + \frac{e}{r_2} = 
1 + \frac{e}{\sqrt{\rho^2 + z^2}} + 
\frac{e}{\sqrt{\rho^2 + (z-a)^2}} \ , \end{equation}

\noindent where $a$ is the Euclidean distance between the two source points, and 
cylindrical coordinates were introduced in the second step. It is not difficult 
to read off the ADM masses in the three external regions, by studying how the 
metric falls off at infinity (using the coordinate $r'_1 = e_1^2/r_1$ as a 
coordinate in the first region, say). One finds \cite{BL}

\begin{equation} M_1 = M_2 = 2e\left( 1 + \frac{e}{a}\right) \ , \hspace{6mm} 
M_3 = 4e \ . \end{equation}

\noindent It is clear that if $a >> e$ the two throats will not affect 
each other much, and there will be only two minimal spheres. But if $a = e$ 
all three masses will be equal, and by symmetry there will be three throats, 
and three minimal spheres. Interestingly the area of the outermost minimal 
sphere---outermost from the point of view of the third external region---is 
smaller than the sum of the areas of the two minimal spheres it is surrounding. 
This will be so whenever the distance $a$ is small enough. There 
must exist a critical value $a_{\rm cr}$ when the third minimal sphere first 
appears. To calculate this we assume 
that any minimal surface is axisymmetric and given by $\rho = \rho (\sigma)$, 
$z = z(\sigma )$, $\phi = \varphi$. The minimal surface equation is then 
obtained by extremising the area functional 

\begin{equation} A = \int \sqrt{\det \gamma}d\sigma d\varphi = 
2\pi \int_{\sigma_0}^{\sigma_1} 
\rho \omega^4\sqrt{\dot{\rho}^2 + \dot{z}^2}d\sigma \ . \end{equation}

\noindent We must find solutions corresponding to closed surfaces. 
Unfortunately---because the stated purpose of this review is to give 
pen and paper examples---numerical methods must be used to do 
this. Moreover the numerical calculation is non-trivial, as is 
clear from the fact that the first attempts to do it did not give quite 
the correct answer. The conclusion eventually turned out to be that 
the critical Euclidean distance is $a_{\rm cr} \approx 1.53 e$ 
\cite{Ca}, and in fact a pair of minimal surfaces are created when we 
go below this value \cite{Bi}. 

Do take note of the fact that the surrounding minimal sphere is there for a 
reason. We assume that cosmic censorship holds, and that we have found the area 
$|S|$ of a---not necessarily connected---trapped surface in our initial 
data. Then there must be an event horizon intersecting the initial 
hypersurface, and since it lies outside the trapped surface we expect its 
area to be larger than $|S|$. In the future we expect the system to settle 
down into a Kerr black hole, and since the area of the event horizon can 
only grow the area of the final event horizon is also larger than $|S|$. 
Some of the initial mass $M$ will be radiated away, so the area of the final 
event horizon---which is never larger than the area of a Schwarzschild 
horizon of the same final mass---will be no larger than $16 \pi M^2$. 
Tracing through this string of inequalities we find that   

\begin{equation} 16\pi M \geq \sqrt{16 \pi |S|} \ . \label{Peninequal} 
\end{equation}

\noindent The mass $M$ and the area $|S|$ are determined by the initial 
data. This interesting strengthening of the positive mass theorem is known 
as the Penrose inequality, and should be a necessary (but not sufficient!) 
condition for cosmic censorship to hold \cite{Penrose}. 
But as the two throats in the Brill-Lindquist initial data are moved closer 
together, the sum of their areas grow \cite{BL}. In fact their sum can 
exceed the bound 
set by the Penrose inequality. What saves the day is precisely the new minimal 
surface that appears out of the blue to surround the two, and now counts 
as outermost \cite{Gibbons}. 

A possible loop hole in the above argument---quite apart from a possible 
failure of cosmic censorship---appears 
in its first step, since it is not clear that a surface that surrounds 
another must have a larger area. However, in the case of time-symmetric data 
the marginally trapped surfaces are minimal and the reasoning does not need 
any amendment, provided the trapped surface is taken to be the (not necessarily 
connected) outermost marginally trapped surface. One can construct simple 
examples in Oppenheimer-Snyder dust collapse where more care is needed 
\cite{Horowitz, IBD}.   

\

{\bf 4. The 2 + 1 dimensional toy model}

\

\noindent To bring the multi-black hole problem within the range of pen 
and paper methods Brill \cite{Dieter}, and Steif \cite{Steif}, eventually 
turned to 2 + 1 dimensions where a trapped ``surface'' is a closed 
spacelike curve with a future pointing timelike Frenet normal vector. There are 
considerable simplifications because there is no Weyl tensor and no shear 
in the Raychaudhuri equation for a null congruence: in 2+1 dimensions

\begin{equation} \dot{\theta} = - \theta^2 - R_{ab}t^at^b \ . \end{equation}

\noindent If we impose Einstein's vacuum equation $R_{ab} = \lambda g_{ab}$ 
the second term vanishes, $\dot{\theta} = - \theta^2$, and hence 

\begin{equation} \theta (0) = 0 \hspace{5mm} \Leftrightarrow \hspace{5mm} 
\theta (\tau ) = 0 \ . \end{equation}

\noindent This has the consequence that a marginally trapped surface must 
lie on a lightlike plane, or---to use a terminology that is more accurate 
in the de Sitter and anti-de Sitter cases---on a lightcone with its vertex 
on \scri . In turn this means that we can easily get a complete 
overview of all marginally (outer) trapped curves in a spacetime of this  
type. Still, even though the vacuum equations imply that spacetime has 
constant curvature, the model is not completely trivial. 

There are no closed trapped 
curves in Minkowski space, but they do exist in a suitable region of Misner 
space, which is Minkowski space with points related by some discrete 
Lorentz boost identified \cite{Hawking}. Since the boost has a line of 
fixed points there is a kind of singularity to the future of such a 
trapped curve, as well as a breakdown of the causal structure in the 
form of closed timelike curves. Misner space does not have a black hole, 
since it has no sensible notion of future \scri, but the same construction 
carried out in anti-de Sitter space has. It gives rise to the BTZ black hole 
\cite{BTZ}. To see how this works consider the metric of 2 + 1 
dimensional anti-de Sitter space, 

\begin{equation} ds^2 = - \left( \frac{1+\rho^2}{1-\rho^2}\right)^2 dt^2 
+ \frac{4}{(1-\rho^2)^2}(d\rho^2 + \rho^2d\phi^2) \ . \label{korv} \end{equation}

\noindent At $t = 0$ we have a time-symmetric initial data slice which in 
these coordinates appears in the guise of the Poincar\'e disk, a space of 
constant negative curvature. The group of isometries preserving this slice 
is isomorphic to the subgroup of M\"obius transformations that preserve 
its conformal boundary (in these coordinates the unit circle). Among them 
are the hyperbolic M\"obius transformations, that have two fixed points on 
the boundary of the disk and a special flow line which is the unique 
geodesic connecting the fixed points---we recall that the geodesics on 
the Poincar\'e disk are arcs of circles or straight lines meeting the 
boundary at right angles. 

\begin{figure}
        \centerline{ \hbox{
                \epsfig{figure=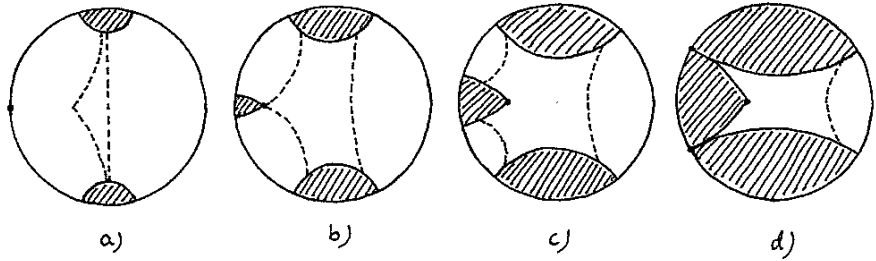,width=120mm}}}
        \caption{\small{ a) The particle comes in from infinity just 
    as a marginally trapped curve is formed in the centre. The event 
    horizon has a kink and does not contain any marginally trapped 
    curves. b) The particle meets the event horizon. c) The event 
    horizon is now smooth and marginally trapped. d) The 
    identification surfaces meet in a singular point at infinity. 
    Reproduced with permission \cite{Emma}.}}
        \label{fig:trappex6}
\end{figure}

Now pick such a M\"obius transformation, obtainable by exponentiating 
an infinitesimal transformation. We identify all the points 
one can connect with it. The result is a space of constant curvature and 
cylindrical topology, having exactly one closed geodesic---a minimal 
``surface''---corresponding to the unique geodesic flow line. These are 
the initial data of the BTZ black hole (and occurs as panel a in 
Fig. \ref{fig:trappex6}). Since anti-de Sitter space is not globally 
hyperbolic these data do not in fact determine the future evolution, but we can 
define the spacetime as the quotient of anti-de Sitter space with the 
discrete isometry group generated by the group element we used to construct 
our initial data. This spacetime is the BTZ black hole \cite{BTZ}. It has a singular 
future (since the isometry will have fixed points to the future, and to 
the past, of the initial data slice), and it will be asymptotically anti-de 
Sitter in the sense that the first order structure of the quotiented 
conformal boundary is conformally isomorphic to the 1+1 Einstein universe, 
the \scri \ of anti-de Sitter space \cite{brapapper}. The event horizon is 
the boundary of the past of \scri , as usual. 

\begin{figure}[ht]
\centerline{ \hbox{
                \epsfig{figure=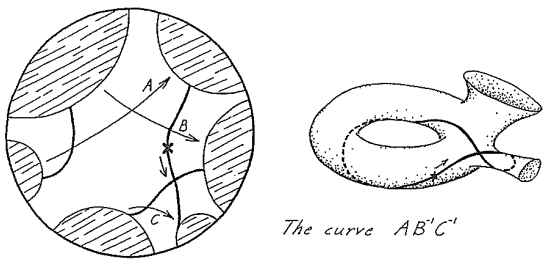,width=80mm}}}
\caption{\small{An initial data surface for a black hole spacetime with two asymptotic 
regions. The white part is a fundamental region. 
The group $\Gamma$ is generated by the group elements $A,B,C$, 
and a closed curve associated to the group element $AB^{-1}C^{-1}$ is shown. 
Reproduced with permission \cite{Holst}.}} 
\label{fig:trappex3}
\end{figure}

This idea can be generalised. Given a pair of geodesics on the 
Poincar\'e disk at $t=0$ there is a unique hyperbolic M\"obius transformation 
taking one to the other. If we choose several such pairs of geodesics---in 
Fig. \ref{fig:trappex3} we choose three pairs---we obtain several group elements. 
If we make 
sure that the geodesics do not intersect these group elements generate a 
free group $\Gamma$, and we define our spacetime as the quotient of anti-de 
Sitter space by $\Gamma$. Its initial data surface at $t = 0$ is guaranteed 
to be a smooth surface of constant negative curvature and one or several 
asymptotic regions. We can now search for closed geodesics---marginally 
trapped surfaces---on this initial data surface. 

But this is not a hard task. Take any closed curve in the quotient space. 
It can be associated to a particular element in the discrete group $\Gamma$, 
as shown by example in Fig. \ref{fig:trappex3}. This group element is in 
itself a hyperbolic 
M\"obius transformation, and is therefore associated with a unique geodesic 
flowline. The latter will become a closed geodesic in the quotient space. 
In this way we find one closed marginally trapped curve for every topologically 
distinct class of closed curves. From a spacetime point of view, and from 
our discussion of the Raychaudhuri equation, we know that a marginally 
trapped curve must lie on a lightcone with its vertex on \scri . This vertex 
can be found by analysing the action of $\Gamma$ on \scri , but since this 
pleasant task has been described elsewhere we do not go into this here 
\cite{Brill}. 

I should add that the restriction to time symmetric spacetimes can be lifted, 
at the expense of rather more work \cite{Holst}.

To get a more lively picture we can try to add some matter to the model. 
So as to not compromise its basic simplicity this is often done in the form 
of ``point particles'', conical singularities with their vertices along 
timelike or null geodesics \cite{DJtH}. There will be a non-trivial holonomy 
associated to any spacelike curve surrounding the singular geodesic, and 
from this one can read off the ``mass'' of the resulting spacetime. Now 
suppose we start with a single BTZ black hole, choose a radial null 
geodesic leaving \scri \ and heading for the event horizon, choose a 
suitable wedge with that geodesic as its edge, and identify the two 
boundaries of the wedge using an element of the isometry group. This then 
describes a lump of matter falling into a black hole, and the question 
arises how it will affect the black hole when it hits \cite{Emma}.   

In any dimension one expects the outermost marginally 
trapped surface to ``jump'' outwards when a lump of matter hits 
it. It will be a somewhat non-local jump since it takes 
place also on that side of the black hole which is not yet in 
causal contact with the infalling matter. What happens there is that some 
locally trapped surface is ``suddenly'' a part of a closed trapped surface 
because of the way the lump of matter curves space elsewhere---while 
it would have been part of an open surface if no matter had been falling in. 
But this is not a spherically symmetric situation, and gravitational 
radiation will prevent any pen and paper calculation from being made---unless 
we are in 2+1 dimensions, where gravitational radiation does not happen.

Concretely, let a massless particle enter a BTZ black hole spacetime at 
$t = 0$ (panel a in Fig. \ref{fig:trappex6}). We follow the time development by slicing 
covering space with a sequence of Poincar\'e disks at increasing values of 
$t$. The particle moves inwards trailing its wedge behind it. Meanwhile the 
geodesic arcs that bound the fundamental region of the BTZ black hole are 
moving closer to each other. Eventually they meet the boundary of the wedge; 
the result is a fixed point on \scri \ that corresponds to a singular
point in the quotient space (which will then continue inwards along a 
spacelike geodesic). The backwards light cone from that singular point 
is the event horizon of the resulting spacetime.

\begin{figure}
        \centerline{ \hbox{
                \epsfig{figure=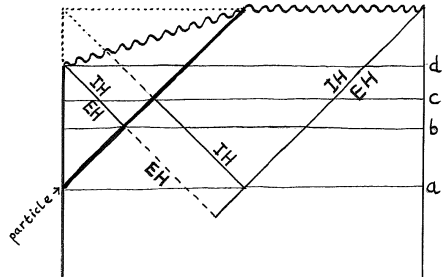,width=70mm}}}
        \caption{\small{ A conformal diagram of a massless particle falling into a 
        BTZ black hole. Four spatial slices corresponding to the panels 
        in Fig \ref{fig:trappex6} are given, and we also see how the isolated 
        horizon (IH) ``jumps''.}}
        \label{fig:trappex4}
\end{figure}

Following the event horizon backwards in time we observe that it is a smooth 
surface foliated by closed marginally trapped curves only as long as it passes 
through the wedge behind the particle (down to panel b in Fig. \ref{fig:trappex6}). 
At earlier times the event horizon has a kink, and does not contain 
any smooth and closed spacelike curves.  Yet there is a closed marginally 
trapped curve at $t = 0$, namely the one that would have evolved into the 
event horizon of the BTZ black hole, had there been no massless particle 
to complicate matters. This will evolve into an isolated horizon which 
is destroyed by its encounter with the particle (in panel c, although only 
the horizon bordering the second asymptotic region is shown in the 
figure). A spacetime picture of all this is given in Fig. \ref{fig:trappex4}.

As a final remark, let us go to 3+1 dimensions by means of the obvious amendment 
of the metric (\ref{korv}), i.e. by adding an extra angular coordinate $\theta$. 
At $t = 0$ we now have a Poincar\'e ball, and there 
are isometries mapping totally geodesic surfaces---spherical caps meeting 
the boundary at right angles---to each other. Take the quotient of 3 + 1 
dimensional anti-de Sitter space with such an isometry. The result is a 
genuine black hole spacetime, but a curious one since its event horizon is 
not a Killing horizon. It grows. One can now look for marginally trapped 
surfaces in a slicing that respects the symmetries of the quotient space, 
and one finds that they foliate a spacelike dynamical horizon situated well 
inside the event 
horizon \cite{Peldan}. The conformal boundary is connected, and has the topology 
of a torus times the real line. This pathological feature apart it is an 
interesting example. 

\

{\bf 5. Collapsing null shells}

\

\noindent A rich supply of interesting trapped surfaces can be had by sending 
a thin shell of incoherent radiation (also known as null dust) into Minkowski 
space. The surfaces will arise as cross sections of the null shell. We want 
the interior of the shell to remain flat, meaning that the shell cannot have 
caustics to the past of the surface. To ensure this we insist that its cross 
section with some spacelike hyperplane is convex. To the future the generators 
of the shell cross, and 
a spacetime singularity results. About the outside of the shell we know 
very little---unless the shell and the mass distribution are spherically 
symmetric there will be gravitational radiation there. If a cosmic censor 
is active a black hole will develop, and eventually settle down to the 
Kerr solution. Again we have the Penrose inequality (\ref{Peninequal}); in 
fact this was the setting in which the inequality was first proposed 
\cite{Penrose}. What is 
remarkable about the shell construction is that both $M$ and $|S|$ can be 
evaluated without knowledge of the exterior of the shell. They result from 
calculations in flat space. 

The energy-momentum tensor of the shell is 

\begin{equation} T^{ab} = 8\pi \mu k_{-}^ak_{-}^b\delta \ , \end{equation}

\noindent where $\mu$ is an arbitrary function on a cross section and the 
delta function is defined with respect to the volume form induced by the 
ingoing null normal $k_-^a$. We now choose the cross section we want by 
intersecting the shell with an outgoing 
null congruence having outer null expansion $\theta_+^{\rm int}$. This 
null expansion will jump at the shell. From the Raychaudhuri equation we obtain 

\begin{equation} \theta_+^{\rm ext} = \theta_+^{\rm int} - 16 \pi \mu \ . 
\end{equation}

\noindent Since the mass distribution $\mu$ is at our disposal we can turn 
any cross section into a marginally trapped surface in this way. Moreover 
we can search for trapped surfaces on the shell---in particular for the 
outermost marginally trapped surface---in an efficient way. 

As an example of the kind of insights one can get from this model, consider 
the case when the shell admits an ellipsoidal cross section. Using a foliation 
of Minkowski space with the usual $t =$ constant hypersurfaces one then 
finds that the caustic appears first at the ends, and so does any trapped 
surface sitting on the shell. The result is that there are no trapped 
surfaces on any $t =$ cross section before the value of $t$ for 
which the singularites first appears \cite{Pelath}. This provides some food 
for thought, given that the numerical algorithms look for marginally trapped 
surfaces in given spacelike hypersurfaces \cite{Thornburg}. (Even in the 
Schwarzschild solution it is known that there exists a foliation with 
spacelike hypersurfaces which reaches all the way to the singularity even 
though its leaves do not contain any trapped surfaces \cite{Iyer}. But that 
foliation was chosen specifically to make this true, while the foliation 
used for the collapsing shell is perfectly natural.)

\begin{figure}
\centerline{ \hbox{
                \epsfig{figure=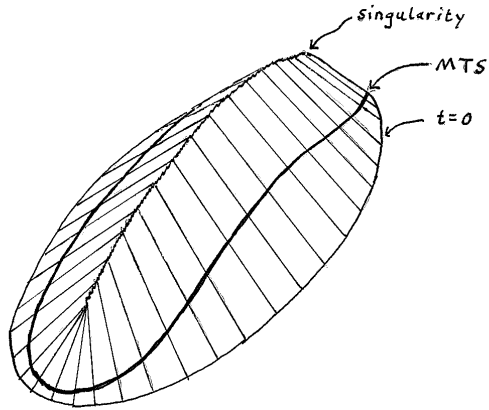,width=80mm}}}
\caption{\small{A collapsing null shell considered by Pelath, Tod and Wald 
\cite{Pelath}: 
It has an ellisoidal cross section at $t=0$. Both the singularity and the 
first marginally trapped surface ``bend'' in time in such a way that the singularity 
appears at an earlier value of $t$ than does the last piece of the marginally 
trapped surface.}} 
\end{figure}

The proof of the Penrose inequality remains partly elusive. It has been proved for 
the case that the null shell is the backwards light cone of a point \cite{Tod}. 
It has also been proved for the case when the cross section is the intersection 
of the shell with a timelike hyperplane in Minkowski space, 
in which case it reduces to the famous Brunn-Minkowski inequality 

\begin{equation} \oint K dS \geq \sqrt{16\pi |S|} \ , \end{equation} 

\noindent which holds for any convex surface in Euclidean space (and $K$ is 
its mean curvature). Gibbons, who was the first to draw this conclusion, therefore 
refers to the Penrose inequality as an isoperimetric inequality for black holes 
\cite{Gary}. The freedom in choosing the trapped surface can be used to test 
other conjectures, notably the hoop conjecture has been studied in this way 
\cite{Pelath}. But this is not the place to pursue these matters. Instead we would 
like to take a peek behind the shell. 

\

{\bf 6. The Vaidya solution}
\

\noindent To have a manageable exterior we make the collapsing shell, and its 
mass distribution, spherically symmetric. We also give the shell a finite 
width. In other words we will look at the Vaidya solution. The matter is still 
an infalling null dust. The metric, using Eddington-Finkelstein coordinates, is

\begin{equation} ds^2 = - \left( 1 - \frac{2m(v)}{r}\right)dv^2 + 2dvdr 
+ r^2d\theta^2 + r^2\sin^2{\theta}d\phi^2 \ . \label{m-metrik} \end{equation}

\noindent Ingoing null geodesics are given by $v =$ constant. The mass 
function $m(v)$ is specified, as a monotone function of the advanced time $v$, 
on \scri$^-$. We will assume that it vanishes for $v < 0$, and becomes constant 
again at some later value of $v$, which means that we have a shell of finite 
thickness, entering Minkowski space and leaving a Schwarzschild black hole 
behind. There are some restrictions on $m(v)$ that must be imposed in order 
to prevent naked singularities---due to the unphysical features of the matter 
model---from occurring. For pragmatic reasons we choose one that enables us 
to solve explicitly for outgoing radial null geodesics, namely 

\begin{equation} m(v) = \left\{ \begin{array}{lll} 0 \ , & v \leq 0 & \ 
\mbox{(Minkowski region)}  \\ \\
\mu v \ , & 0 \leq v \leq M/\mu & \ \mbox{(Vaidya region)} \\ \\  
M \ , & v \geq M/\mu & \ \mbox{(Schwarzschild region ) \ . } 
\end{array} \right. \label{mu} \end{equation}  
 
\noindent When presented in this way the spacetime is not $C^1$, but this 
can be mended. The constant $\mu$ must be set larger than $1/16$ to avoid naked 
singularities \cite{Papa}. The linearly rising mass function has the special 
feature that the Vaidya region admits a homothetic Killing vector 

\begin{equation} \eta = v\partial_v + r\partial_r \ , \end{equation}

\noindent whose flow lines lie in hypersurfaces with constant $x = v/r$. 
This is a simplifying feature since a trapped surface remains trapped if 
moved by a homothety \cite{Carrasco}. 

Now where are the trapped surfaces? The first observation is that the 
hypersurface 

\begin{equation} r = 2m(v) \end{equation}

\noindent is foliated by round and marginally trapped spheres. In the 
Schwarzschild region it is the event horizon, but in the Vaidya 
region it is a spacelike dynamical horizon lying well inside the event 
horizon. We will refer to it as the ``apparent 3-horizon'', because it 
needs a name. Cosmic censorship requires that all trapped surfaces are 
confined to the interior of the event horizon \cite{Hawking}. In the Schwarzschild 
case this is easily proved directly, because a trapped surface cannot 
have a minimum in $t$, where $t$ is the parameter along a hypersurface 
forming timelike Killing vector field \cite{MS}. In the Vaidya region 
there is a variation of this argument employing the Kodama vector 
field---which gives the direction in which the area of the round spheres 
is unchanged---leading to the conclusion that any trapped surface must 
lie at least partly inside the apparent 3-horizon. It may extend partly 
outside, but there will be a special spacelike hypersurface (of constant 
``Kodama time'') in its exterior which the trapped surfaces cannot reach 
\cite{BenSen}. See Senovilla's chapter in this book. 

The introduction done with, let us look for some examples. 
We begin by looking for ``tongues'' sticking out of the apparent 3-horizon. 
A naive way to do so is to begin by visualising the solution in the 
$(v, r,\theta)$-coordinates, suppressing one coordinate that we will not 
use. The apparent 3-horizon then forms a cone with its vertex at the origin, 
which eventually joins the cylinder representing the Schwarzschild part of 
the event horizon. We introduce another cone meeting the apparent 3-horizon 
in a marginally trapped round sphere (a circle in the picture), and look at 
2-surfaces that are cross sections of that second cone. In equations 

\begin{equation} v = \frac{k}{2\mu}r - v_0 \ , \hspace{8mm} v = \frac{1}{2\mu}
r + a(\theta) \ . \end{equation}

\noindent When $a = 0$ we are on the apparent 3-horizon, and wherever $\theta$ 
is such that $a(\theta) 
< 0$ the tongue is sticking out if it. There are some restrictions that must 
be imposed if it is to do so, for instance that $k > 0$ \cite{Ashtekar, BenSen}. 
Simple choices are 

\begin{equation} k = 3 \ , \hspace{8mm} a(\theta ) = a_0 + a_1\cos{\theta} \ . 
\end{equation}

\noindent To first order in a perturbation expansion one finds that the tongues are 
trapped surfaces if and only if $a_0 > 0$. By choosing $a_1$ suitably we can 
clearly arrange that they stick partly outside the apparent 3-horizon. But 
how far out can they go?

\begin{figure}
\centerline{ \hbox{
                \epsfig{figure=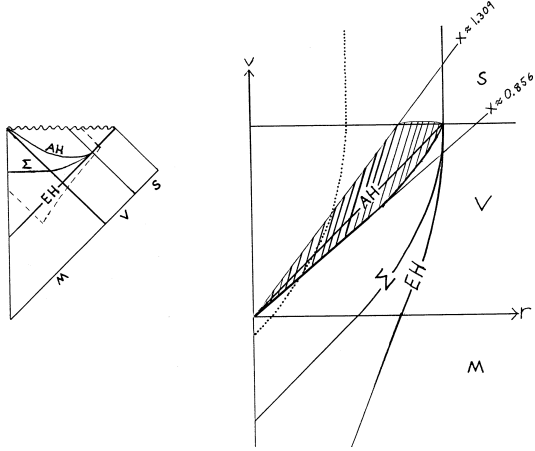,width=70mm}}}
\caption{\small{A Penrose diagram of a Vaidya spacetime, and a $v-r$ diagram of 
the region inside the dashed curve. It includes a part of the event horizon (EH), 
the spacelike part of the apparent 3-horizon (AH), and a hypersurface $\Sigma$ 
below which no trapped surfaces can extend \cite{BenSen}. The tongues discussed 
in the text are 
confined between two lines of constant $v/r$. There are also trapped surfaces 
to the left of the dotted curve.}} 
\label{fig:ERE09}
\end{figure}

The inequality that guarantees that the tongues are trapped is an unwieldy one, 
even in this simple case, and we had to rely partly on Mathematica in order to 
analyse it. For the case $\mu = 1/2$ the answer is that the tongues remain trapped 
provided that 

\begin{equation} 0.856 < x = \frac{v}{r} < 1.309 \ . \end{equation}

\noindent This is how far a maximally extended trapped tongue, of this 
particular kind, extends---provided 
it stays in the Vaidya region, where the result in fact reflects the behaviour 
of trapped surfaces under homotheties \cite{Carrasco}. If a tongue hits 
the Schwarzschild region there are further restrictions; see Fig. \ref{fig:ERE09} 
for the final result, and \AA man et al. \cite{ABS} 
for other choices of $\mu$ and $a({\theta})$. The Gaussian curvature of this 
tongue is positive, and maximal at its ``tips''. Its area grows as it extends 
more, at least to second order in a perturbation around a round sphere.

Of course this is a very naive construction. It would be more interesting to 
produce a marginally trapped tongue of this general type. Using general 
theorems \cite{Lars, Andersson} one can prove \cite{BenSen} that 
they exist, and that they ``evolve'' into marginally trapped tubes intersecting 
the apparent 3-horizon, but to get them in explicit form seems to be very
hard.

Can the trapped surfaces extend all the way into the flat region? Actually 
for sufficiently large $\mu$ the tongues we just discussed can do that, but 
let us try to design a surface that passes through the centre, 
also for $\mu = 1/2$. We begin with 
the observation that locally trapped surfaces certainly exist in Minkowski 
space. A flat 2-plane has both its null expansions vanishing. So suppose we 
define a surface by setting 

\begin{equation} \theta = \frac{\pi}{2} \ , \hspace{10mm} v = v(r) \ . 
\end{equation}

\noindent In the flat region this is marginally trapped if $t = v-r =$ 
constant. The idea is to ``steer'' it through the Vaidya region in such 
a way that it remains marginally trapped there, and see what it looks 
like when it emerges into the Schwarzschild region.

With our simple choice for the mass function it is easy to show that the 
surface remains minimal throughout the Vaidya region if we let $v(r)$ be 
a solution of the differential equation 

\begin{equation} \frac{dv}{dr} = \frac{1}{1-\frac{\mu v}{r}} \ . \end{equation}

\noindent We join this surface to the flat 2-plane in the flat region and find 
that its differentiability, in these coordinates, 
is $C^{2-}$. On the boundary to the 
Schwarzschild region it forms a round circle. Now it is interesting to 
observe that the cylindrical Schwarzschild surface 

\begin{equation} \theta = \frac{\pi}{2} \ , \hspace{8mm} r = M \ , 
\end{equation}

\noindent has both its null expansions vanishing, and it can be joined to 
one of the surfaces that we have steered through the Vaidya region by 
a suitable choice of integration constants. The result is a surface 
which is minimal throughout the entire spacetime---but its topology is 
that of a sphere with a point removed ``at'' the singularity, so this 
is not a closed marginally trapped surface.

But we can wiggle the surface a little bit, so that it becomes locally trapped, 
and so that it arrives to the Schwarzschild region with negative slope in the 
$v-r-$diagram. Continuing into Schwarzschild with fixed slope the null 
expansions become increasingly negative, and once they are sufficiently 
negative we can close off the surface with a spherical cap---we have constructed 
a closed trapped surface that extends into the flat part of the Vaidya 
spacetime. (Full details, for almost the same construction, can be found 
elsewhere \cite{BS}.) 

On the one hand this looks odd---one might have thought that no closed 
trapped surfaces could enter the flat region. But indeed in order to 
see that they are closed one has to collect information from a much larger 
region of spacetime. From another point of view it is somehow satisfactory, 
because it means that one cannot reach the singularity without crossing a 
trapped surface. Actually one can prove \cite{BenSen} that it is impossible for 
a trapped surface to enter the flat region if $\mu \leq 1/8$. But the smaller 
one makes $\mu$ the closer one gets to a nakedly singular spacetime, so that 
is presumably at the unphysical end. 

A more systematic search for marginally trapped surfaces in the Vaidya 
solution, using a different mass function, an axisymmetric slicing, and 
a horizon finder, has been made by Nielsen et al. \cite{Nielsen}. A more 
systematic study of spherically symmetric spacetimes in general will be 
found in Senovilla's chapter in this book.

If we are content with outer trapped surfaces a different picture emerges. 
Then a trapped surface in the deep interior of the Schwarzschild region can 
develop a ``tendril'' reaching down through the Vaidya region and indeed 
through any point in the interior of the black hole, also in the flat 
diamond \cite{Ben-Dov}. Once it emerges into the flat region the tendril lies in a tubular 
neighbourhood of a (null) generator of the event horizon. 
Casual inspection in the 
flat region would suggest that the tendril is part of an untrapped surface 
with its inner expansion negative, but in fact when the global structure of 
the spacelike slice is examined one sees that it is outer trapped. 
This behaviour was suggested by Eardley, who in fact conjectured that the 
boundary of the region through which outer trapped surfaces pass is always 
the event horizon, in any black hole \cite{Eardley}. 

\begin{figure}
\centerline{ \hbox{
                \epsfig{figure=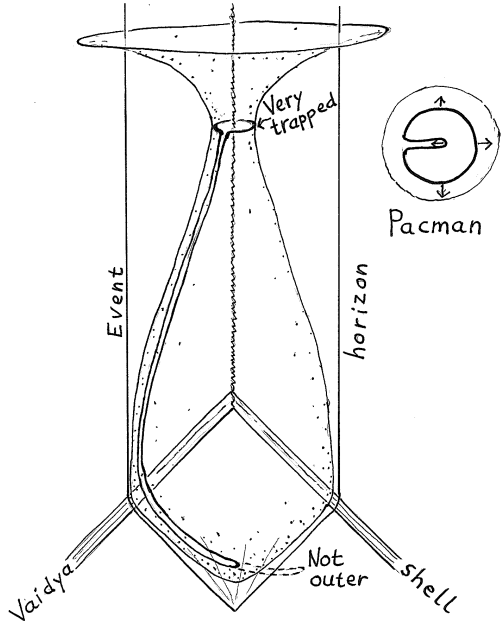,width=70mm}}}
\caption{\small{An impressionistic picture---$v$-$r$-coordinates would 
not be so useful here---of Ben-Dov's construction \cite{Ben-Dov}. 
A spacelike hypersurface 
sits inside the Vaidya event horizon, and it contains a very trapped 
surface around its ``neck'', close to the singularity. This surface has 
developed a ``tendril'' reaching down to Minkowski space. It remains outer 
trapped, as one can see by flattening the spacelike hypersurface (at right). 
As far as the intrinsic properties of the tendril are concerned, they remain 
the same if the tendril extends outside the event horizon (shown dashed), 
but then no spacelike hypersurface can contain both the tip of the tendril 
and the very trapped surface from which it came.}}
\label{fig:trappex7}
\end{figure}

What prevents the tendril from following its generator 
even further, out of the event horizon? The answer is that should it do so, 
it would no longer be lying in a suitable spacelike hypersurface, and would 
not count as ``outer'' any more. Indeed we assume it to be embedded 
in a spacelike hypersurfaces containing the trapped surface from which the 
tendril emerged in the first place. See Fig. \ref{fig:trappex7}. 
The lesson is that we can decide whether 
a surface is trapped by inspection of the surface itself, while this is 
not possible for outer trapped surfaces. 

\

{\bf 7. Envoi}

\

\noindent We hope that the reader enjoyed these examples. If not, the list 
of references will provide more satisfying reading. Meanwhile the examples 
may perhaps serve to drive home the point that being trapped or outer trapped 
is a spatially non-local property of a surface---and also that the 
condition for being outer trapped 
is substantially easier to fulfill. 

\

{\bf 8: Acknowledgments}

\

\noindent I thank Jos\'e Senovilla for a 
very enjoyable collaboration, Emma 
Jakobsson for permission to include some results from her Master's thesis, 
and the Swedish Research Council for support under contract VR 621-2010-4060.

\end{document}